%% file: journal_smartgrid.tex
\documentclass[journal]{IEEEtran}
\usepackage{graphicx}
\usepackage{xcolor}

\usepackage{multirow}
\usepackage{multicol}
\usepackage{lipsum}
\usepackage{graphicx}
\usepackage{graphics}
\usepackage{color, colortbl}
\usepackage{array}
\usepackage{siunitx}
\usepackage{amsmath}
\usepackage{amsbsy}
\usepackage{float}
\usepackage{caption}
\usepackage{subfigure}
\usepackage{cleveref}
\usepackage{tcolorbox}
\usepackage{amssymb}
\usepackage{scalerel}
\usepackage{mathabx}
\usepackage{verbatim}
\usepackage{booktabs}
\usepackage{rotating,tabularx}
\usepackage{mathrsfs} 
\usepackage{url}
\usepackage{adjustbox}
\usepackage{soul}
\usepackage{cite}
\usepackage{tikz}
\usepackage{algorithm}
\usepackage{algpseudocode}
\usepackage{color, colortbl}
\usepackage{algorithm}
\usepackage{algpseudocode}
\usepackage{amsmath}
\usepackage{array}
\usepackage{mdwmath}
\usepackage{mdwtab}
\usepackage{eqparbox}
\usepackage[utf8]{inputenc}
\usepackage[english]{babel}

\usepackage{tikz}  
  
\usetikzlibrary{shapes, arrows, calc, arrows.meta, fit, positioning} 

\tikzset{block/.style={draw, thick, text width=2cm , minimum height=1.3cm, align=center},   
line/.style={-latex}     
}

\tikzset{  
    -Latex,auto,node distance =1.5 cm and 1.3 cm, thick,
    state/.style ={ellipse, draw, minimum width = 0.9 cm}, 
    point/.style = {circle, draw, inner sep=0.18cm, fill, node contents={}},  
    bidirected/.style={Latex-Latex,dashed}, 
    el/.style = {inner sep=2.5pt, align=right, sloped}  
}

\newcommand*\ttvar[1]{\textit{\expandafter\dottvar\detokenize{#1}\relax}}
\newcommand*\dottvar[1]{\ifx\relax#1\else
  \expandafter\ifx\string_#1\allowbreak\else#1\fi
  \expandafter\dottvar\fi}
\newcommand\longvar[1]{\mathchardef\UrlBreakPenalty=100
\mathchardef\UrlBigBreakPenalty=100\url{#1}}

\ifCLASSINFOpdf
\else
\fi
\begin{document}
%
\title{
A Firewall Optimization 
for 
Threat-Resilient 
Micro-Segmentation in Power System Networks
%
}
%
%
%

\author{
        Abhijeet~Sahu,~\IEEEmembership{Student Member,~IEEE,}
        Patrick~Wlazlo,~\IEEEmembership{Student Member,~IEEE,}
        Nastassja~Gaudet,
        Ana~Goulart,~\IEEEmembership{Member,~IEEE,}
        Edmond~Rogers,
        and~Katherine~Davis,~\IEEEmembership{Senior Member,~IEEE}
\thanks{A. Sahu, N. Gaudet, A. Goulart, K. Davis are with Texas A\&M University, College Station, TX. P. Wlazlo is with Vistra Energy. Edmond Rogers is with University of Illinois, Urbana Champaign. 
}
\thanks{}
}

%
%

\markboth{IEEE ,
September 2023}%
{Shell \MakeLowercase{\textit{et al.}}: Bare Demo of IEEEtran.cls for IEEE Journals}
%



\maketitle

\begin{abstract}
Electric power delivery relies on a communications backbone that must be secure. 
SCADA
systems are essential to critical grid functions and include
industrial control systems (ICS) protocols such as the Distributed Network Protocol-3 (DNP3). 
These protocols are vulnerable to cyber threats that power systems, as cyber-physical critical infrastructure, must be protected against.
For this reason, the 
NERC
Critical Infrastructure Protection standard CIP-005-5 specifies that an electronic system perimeter is needed, accomplished with firewalls. 
%
This paper presents how these electronic system perimeters can be optimally found and generated using a 
meta-heuristic approach for optimal security zone formation for large-scale power systems. 
Then, to implement the optimal firewall rules in a large scale power system model,
this work presents a prototype software tool that takes the optimization results and auto-configures the firewall nodes for different utilities in 
 a cyber-physical testbed. Using this tool, 
 firewall policies are configured for all the utilities and their substations within 
a synthetic 2000-bus model, assuming two different network topologies. 
Results generate the optimal 
electronic security perimeters to protect a power system's 
data flows and compare the number of firewalls, monetary cost, and risk alerts from path analysis.
\end{abstract}


%
\IEEEpeerreviewmaketitle
\input{Frontiers_Latex_Templates/Journal_SectionWise/1_Introduction}
\input{Frontiers_Latex_Templates/Journal_SectionWise/2_Background}

\input{Frontiers_Latex_Templates/Journal_SectionWise/3_Problem_Formulation}
\input{Frontiers_Latex_Templates/Journal_SectionWise/4_Proposed_Solution}

\input{Frontiers_Latex_Templates/Journal_SectionWise/5_Abridged_Tool}

\input{Frontiers_Latex_Templates/Journal_SectionWise/6_Results_Analysis}
\input{Frontiers_Latex_Templates/Journal_SectionWise/7_Discussion}
\input{Frontiers_Latex_Templates/Journal_SectionWise/8_Conclusion}

\section*{Acknowledgments}
This research is supported by the US Department of Energy's (DOE) Cybersecurity for Energy Delivery Systems program under award DE-OE0000895.

\bibliographystyle{IEEEtran}
\bibliography{myReferencev2}

\end{document}

%% file: Frontiers_Latex_Templates/Journal_SectionWise/1_Introduction.tex
\section{Introduction}\label{intro}

Cybersecurity is crucial
to ICS protocols because a cyber threat can have disastrous consequences~\cite{ukraine}. It is important to configure the communications network to protect field devices, utility control centers (UCC), and balancing authorities (BA) from as many threats as possible. Current defenses are generally based on the defense-in-depth approach as well as industry guidelines and requirements, especially the North American Electric Reliability Corporation (NERC) Critical Infrastructure Protection (CIP) standards.
%
These are existing tools and techniques
that are being used or considered to protect power systems against cyber threats:

\begin{itemize}
\item \textbf{Firewalls} are deployed to inspect and filter all network traffic coming in and out of a network, following rules that allow or deny certain data packets.  Firewalls, or an \emph{electronic system perimeter}, are covered in NERC CIP-005-5~\cite{nerc_cip_005} and
%
carefully configured using access control lists (ACLs).
They allow network operators to specify which hosts and application layer protocols are blocked or allowed into their network. 
    \item \textbf{Intrusion Detection Systems (IDS)} such as Snort, Zeek, Suricatta, and other third-party Security Information and Event Management (SIEM) 
    such as Splunk 
    are deployed
    to alert
    a network administrator of a cyber intrusion. Many works investigate IDS, such as to develop
    pre-processors for Distributed Network Protocol-3 (DNP3)
    ~\cite{dnp3_bro}, to adopt 
    the signature-based IDS Snort for the ICS protocol Modbus~\cite{snort_modbus}, and to propose anomaly-based IDS using machine-learning~\cite{ids_ref} 
    to identify zero-day intrusions. 
    
\end{itemize}

Intrusion detection can 
alert a network administrator once an intrusion occurs; this is reactive. By contrast, firewalls block and filter network traffic to protect the system proactively.
%
This paper presents a proactive optimal defense 
to
secure the Operational Technology (OT) networks of electric power utilities through optimized firewall deployment and configuration that includes physical system risk.
The motivation is that a utility's
network, 
particularly firewalls, are an
available 
major resource 
whose value can be maximized with the use of physics and operational risk 
for proactive cyberattack defense.
%
Being able to optimize and automate how these firewalls are generated can help utilities prevent attacks and avoid manual input and associated time delays and errors.


Design of the security perimeters (firewalls) while ensuring
compliance to NERC CIP standards is elaborated in~\cite{TPECFW2022}, that develops a solution and tool for the automatic generation of firewall rules in 
large-scale power systems, and configures
the communications network and simulates the utility dataflows based on the network topology~\cite{cybermodel} of a 2000-bus synthetic grid~\cite{synthetic}.
This paper extends~\cite{TPECFW2022}
by addressing the 
optimization question: \textbf{\textit{What is the best approach to automatically configure the network devices that implement electronic security perimeters in large-scale electric power systems?}}  We propose and validate a meta-heuristic optimization approach that allows stakeholders to examine the tradeoffs in different
criteria and 
metrics
including
(1) security, (2) reliability, and (3) cost. 
Results are presented
using the automated firewall generation tool 
within our cyber-physical \textit{Resilient Energy Systems Lab (RESLab)} testbed.
%
This work uses the cyber model as the foundation and follows~\cite{mycqrpaper} which demonstrates how to perform the NERC-compliant configuration of firewalls in electrical utility companies, from the substation, to utility control center (UCC), and balancing authority (BA), including demilitarized zones (DMZs), by adopting and automating those principles within the optimization technique and automatic firewall generation tool, and demonstrating it in \textit{RESLab}~\cite{testbed_architecture}. 
%
In this paper, we also consider
a realistic mesh network topology for the synthetic communication network 
for identifying optimal security zones, with the objective to
minimize risk and budget of the SCADA system.  
These are this paper's contributions:

\begin{itemize}
    
    \item A software application is introduced that automates the configuration of firewall rules based on NERC CIP and a utility's security policies, demonstrated in the 2000-bus synthetic cyber-physical model. The tool makes automatic firewall rule generation based on industry specifications 
    more scalable and less error prone.
    \item A novel meta-heuristic approach method is proposed for optimal security zone formation for the firewalls deployed in the UCCs  and  their  substations, with the objective to minimize cyber insurance cost and maximize grid resilience.  The optimization approach is implemented and 
    evaluated 
    in the new tool mentioned above.
     \item The 
     firewall rule automated configurations 
     are integrated and evaluated in $RESLab$~\cite{testbed_architecture} as a proposed application in a next-generation cyber-physical energy management system.
\end{itemize}

The rest of this paper is organized as follows. Section~\ref{background} briefly reviews the power system network model used in this work
as well as prior work on network segmentation and security zone creation. Section~\ref{problem} 
formulates the optimal 
security zones to reduce risk to power system resilience as well as investment cost for a utility 
as a multi-objective optimization problem.
The proposed solution based on a Non-dominated Sorting Genetic Algorithm (NSGA) 
is presented in Section~\ref{solution} to solve the optimization.  
Section~\ref{tool} introduces the large scale firewall generation and auto-configuration tool.  
Detailed definition of the problem along with the result analysis 
is covered in Section~\ref{results}. Finally, Section~\ref{discussion} reviews 
this paper's contributions 
and discusses future work.  

%% file: Frontiers_Latex_Templates/Journal_SectionWise/2_Background.tex
\section{Grid Cybersecurity Preliminaries}\label{background}


\subsection{Power Grid Communication Network Overview}
\label{com_model}

The following system preliminaries are used to design our optimization and automatic firewall generation solution. The models are implemented in emulation in RESLab where we validate the performance of the proposed approach through conducting experiments.

\subsubsection{Basic Network Topology}
\label{com_model}
The power system network and its dataflows 
are detailed in~\cite{mycqrpaper,cybermodel}. The 
basic topology includes a substation, a UCC, and a BA, derived from the cyber topology defined in~\cite{cybermodel}. The UCCs are created by clustering substations in the synthetic 2000-bus model and connecting them in a star topology to a UCC. One BA is added to emulate a real BA and connected to all UCCs in a star topology. Initially, all substations are connected to the UCC in a star topology. Later in this paper, we address other network topologies as well. 



\subsubsection{Data/Service Flows}
\label{dataflows_sec}
For optimal firewall design, the following main data/service flows are considered; more details on these commonly used smart grid protocols are provided in~\cite{TPECFW2022}.

\begin{itemize}
\item{DNP3 Protocol:}
The Distributed Network Protocol Version 3 (DNP3) for SCADA uses TCP port 20000. 

\item{Web-based Protocols:}
Hypertext Transfer Protocol (HTTP) and secure HTTP (HTTPS) are used by the Human Machine Interface (HMI) nodes as well as vendor applications through 
a utility's public demilitarized zone (DMZ).

\item{Remote Access Protocols:}
Secure Shell (SSH), using TCP port 22, and Remote Desktop (RDP), using TCP port 3389, allow vendors to remotely access the utility through the vendor DMZ. 

\item{Database Protocol:}
Structured Query Language (SQL) is 
typically used 
to upload or retrieve data from 
historian servers. The TCP port number for SQL is 1433. 

\item{ICCP Protocol:}
The Inter-Control Center Communications Protocol (ICCP) is used by the BA to connect to several utilities with
TCP port number 102. 

\end{itemize}

Incorporating these data flows in the firewall optimal design and emulating them in the $RESLab$ testbed provides a platform to develop data-dependent IDS solutions that improve the defense against a wide variety of threats. Examples include
network-targeted attacks and steps of larger campaigns that include probing, web-based, user-to-root, and remote-to-local attacks. For instance, monitoring the headers and payloads in a HTTP packet, either generated at a local web-server at a substation or a public web-server at a UCC, can be prevented. Web-based attacks like cross-site scripting, SQL injection, parameter manipulation in Web-API request, etc., can be prevented. 

\subsubsection{Firewall Configuration}
\label{firewall_config}
 A typical firewall configuration includes the 
\textit{Interface Configuration} that defines the Electronic Security Perimeter (ESP) it protects, \textit{Object-Groups} based on the services and the host allocation to networks, and \textit{Access-Control Lists}, which are the policies. The 
configurations
are 
elaborated in~\cite{TPECFW2022}, with 
a list of each object group along with the application port numbers and protocols used. 

\subsubsection{Security Zones}
A \textit{security zone} is a portion of a network that has specific security requirements. For instance, consider the example in Fig.~\ref{fig:zonation}, with Utiliy Control Center $UCC_7$ managing four substations named $A, B, C, D$. First, to analyze the scenario without network clustering, the number of security zones for \textit{Substation $A$}, based on
Fig.~\ref{fig:sub_diag}, would be two, one for the $Substation\;DMZ$ and another for the $OT$ side, with station level, bay level, and process level devices.
This needs at least one firewall. For \textit{Substation $B$} in Fig. Fig.~\ref{fig:zonation}, there will be 4 security zones, 1 for the $Substation\;DMZ$ and 3 for the $OT$ (itself and Substation $C$ and $D$), while there are 2 and 3 security zones for Substation $C$ and $D$ respectively. With network clustering as shown in 
Fig.~\ref{fig:zonation}
we propose incorporation of only 6 security zones, i.e., 1 $OT$ and 1 $DMZ$ for each cluster. With clustering, the number of firewalls to be deployed as well as the number of zone-based ACLs to be configured within each firewall reduces. This clustering approach would require an upgrade of routing policies as well as use of VLAN trunking in the network switches. For instance, in the cluster that comprises \textit{Substation $B$} and $C$, suppose that the load relays are part of a single $VLAN$ and hence communication between them would require VLAN trunking to be enabled in the network switch.

\begin{figure}\label{fig:zonation}
\centering
\begin{tikzpicture}  
  [scale=.9,auto=center,every node/.style={circle,fill=blue!20}] 
    
  \node (a1) at (0,2) {A};  
  \node (a2) at (2,5)  {UCC-7}; 
  \node (a3) at (3,2)  {B};  
  \node (a4) at (4,1)  {C};  
  \node (a5) at (4.5,3.5)  {D};  
  \node[circle,draw,minimum size=1.5cm,fill=white,fill opacity=0.1] (c) at (0,2){}; 
  \node[circle,draw,minimum size=1.5cm,fill=white,fill opacity=0.1] (c) at (4.5,3.5){};

  \path[bidirected] (a1) edge (a2); 
  \path[bidirected] (a2) edge (a3);
  \path[bidirected] (a2) edge (a5);
  \path[bidirected] (a3) edge (a5);
  \path[bidirected] (a3) edge (a4);
  

  \path[-]  (4.5,2) edge (4.5,0);
\path[-]  (4.5,0) edge (2,2);
\path[-]  (2,2) edge (3.25,3.25);
\path[-]  (3.25,3.25) edge (4.5,2);
\end{tikzpicture}  
\caption{A sample network cluster with utility control center $UCC_7$ and
\textit{Substations A}, $B$, $C$, and $D$, with $B$ and $C$ in one cluster.}
\label{fig:zonation}
\end{figure}
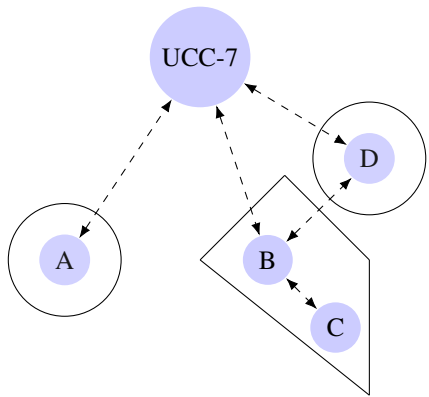

\subsubsection{Firewall and Security Zone Implementation} To implement our firewalls and security zones, we use Cisco Adaptive Security Appliances (ASAs), commonly adopted in industry, and used in other research papers~\cite{case_studies}. The ASA firewalls have a special configuration called a security zone (described above). Each network interface can be set to have a security level from 0-100 that determines its security hierarchy in the network. Incoming packets cannot pass from lower security zones to higher security zones without special rules. 

\subsection{Utility Firewall Optimization Related Work}
\label{relatedwork}
Previous works in optimal firewall policy design have not quantitatively considered grid resiliency.
%
Most works
target improving firewall performance, rather than prioritizing firewall rules based on a quantitative measure of
\textit{importance} of the security perimeter they protect, which is the focus and contribution of our work. 
%
Specifically, the firewalls in a utility network need to consider which specific substation components are regulated from a specific OT network, and the policies need to be ranked
depending on the criticality of the physical components in the electric grid.




The Information Technology (IT) domain has extensively studied optimal firewall policy design, with the main focus on
using firewalls' computing resources 
efficiently. 
A merge model 
to reduce rule filtering time by addressing conflicts with an action constraint strategy is proposed in~\cite{opt_fw1}. Authors in \cite{opt_fw2} develop an automatic discovery of anomalies in firewall policies 
for large-scale enterprise networks. A firewall decision tree is proposed in~\cite{opt_fw3} to identify 
redundant rules and improve
performance. FIREMAN~\cite{opt_fw4} is a static analysis toolkit that models firewall rules using binary decision diagrams and is evaluated in an enterprise network. A filter selection optimization 
is formulated in~\cite{opt_fw5} to minimize 
traffic blocking time while considering costs.
Another rule optimization application~\cite{opt_fw6} removes anomalies in Linux \texttt{iptables} and merges similar rules to improve performance.


Other works study the configuration of firewalls in a utility's SCADA system, such as cases studies in~\cite{case_studies}, 
that 
consider American National Standards Institute (ANSI)/International Society for Automation (ISA) best practices but do not include details on the variety of application protocols or services that need to be configured. The widely used NP-View tool by Network Perception helps facilitate this type of analysis for stakeholders~\cite{npview}. 

The optimal security zone segmentation problem is less studied. For instance,  SONICS~\cite{sz1} Segmentation On iNtegrated ICS systems is a
segmentation method proposed to simplify security zones 
based on 
systems' characteristics 
rather than organizational requirements.
%
Micro-segmentation is a new approach that segregates physical networks into isolated logical networks to limit an adversary's ability to move laterally through network.
Authors in~\cite{sz2} propose an analytical graph-feature based framework to quantify the robustness of the micro-segmentation for enhancing security. Unlike its framework for evaluation, this paper presents a novel method for segmentation that explicitly considers grid resilience post-segmentation. Similarly,~\cite{sz5} proposes 
a 
network segmentation method that is 
based on simulated annealing search and agent-based simulation.

Partitioning an ICS network and placement of firewalls to improve latency for time critical communications is evaluated in~\cite{sz3}. Challenges with this approach include that every ICS protocol has a different payload size, resulting in a different transmission time, and that substations can
vary substantially in their usage of 
ICS protocols.
By comparison, our security zone segmentation targets distribution of critical transmission lines uniformly across multiple security segments based on their Line Outage Distribution Factor (LODF).  
LODFs may be computed using the Injection Shift Factors and
may be iteratively evaluated when more than one line outage is considered. The prominent role of cascading outages in recent blackouts has created a need in security applications to evaluate LODFs under multiple-line outages. 

Prior work in~\cite{bryan_sandia} 
considers network segmentation for reducing load shedding through a tri-level optimization model~\cite{bryan_sandia} on optimal network segmentation through subnetting and creation of virtual local area networks (VLANs). 
In~\cite{bryan_sandia}, network segmentation, through an interdiction Mixed Integer Nonlinear Problem (MINLP) is formulated and solved using bilevel branch and bound, where operations at three levels are optimized, i.e., an IT administrator is given an allowance for network segmentation, an adversary is given a fixed budget to attack the segmented network to inflict damage; in the third level, the grid operator is allowed to redispatch the grid after the attack to minimize load shedding.
%
%
However, challenges arise in 
traditional 
optimization with 
incorporating cyber and physical system objectives and constraints in such a formulation which makes
the solution challenging to obtain and difficult to scale.
Hence, in the current work, a meta-heuristic approach with multi-objective optimization for optimal security zone segmentation is proposed that balances cost and risk, where cost refers to the firewall resources and risk refers to the physical devices that will be under threat.





%% file: Frontiers_Latex_Templates/Journal_SectionWise/3_Problem_Formulation.tex
\section{Optimal Network Clustering Problem}\label{problem}
In this section, we formulate
a method to find optimal network clusters considering hybrid utility network topologies where the resulting 
electronic security perimeters can
reduce cyber investment 
as well as maintain grid resilience. %
%
This paper's proposed solution is to optimally design and deploy firewalls 
by modifying 
their configurations and 
security zones without altering the underlying cyber network topology
to protect an electric grid from cyber intrusions, where the solution's impact 
is based on quantifiable cyber and physical metrics.

\subsection{Components Description}

The following components (both cyber and physical) are
included in the optimal firewall policy design problem.

\subsubsection{Firewall Policy Components}
     \textit{(a) Interfaces ($N_{ifc}$):} The number of interfaces that are configured depends on the number of physical interfaces or ports, either serial or Ethernet, 
     to the firewall device.
    \textit{(b) Object Groups ($N_{og}$):} The number of object groups defined in a firewall depends on the number of network and port objects. The port object depends on the number of protocols or services allowed to ingress or egress out of the security zones associated with the firewall, while the number of network object groups depends on the number of security zones in a substation or UCC network. Depending on the degree
    of a 
    node in the graph model, the number of network objects will vary. 
    \textit{(c) Access Control Lists ($N_{acl}$):} The number of ACLs depends on the number of port and network object groups. 

\subsubsection{Physical Device Components}

    \textit{(a) Isolators ($N_{iso}$):} The number of isolators controlled by the relays within a network cluster. An isolator or disconnect
    is a mechanical switch 
    used to isolate a portion of a substation when a fault occurs. Usually they are operated manually
    when the 
    circuit is under a no-load condition. The usual current withstanding capability of the isolators is lower in comparison to the circuit breaker. 
    \textit{(b) Circuit Breakers ($N_{cb}$):} The number of circuit breakers controlled by the relays within a network cluster. Circuit breakers 
    protect the electrical equipment like generators from overload and short circuits
    during operation. 
    \textit{(c) Transmission Lines ($N_{xline}$):} The number of the other controllable elements in the transmission lines apart from isolators, circuit breakers and transformers within a cluster. 
    \textit{(d) Transformers ($N_{xfmr}$):} The number of 
     transformers associated with the substations in a cluster. Transformers are crucial widespread power system components that change voltage levels without altering the frequency.

\subsection{Problem Formulation}
The optimization problem is formulated where the \textbf{decision variables} are the edges in the sub-graph that need to be removed 
to form the network clusters.
The above 
four physical components are given weights $w_{iso}$, $w_{cb}$, $w_{xline}$, and $w_{xfmr}$, respectively.

The following \textbf{objectives} are considered, where \textbf{F1} is the number of firewalls, \textbf{F2} is the number of ACLs, \textbf{F3} is a physical security metric, and \textbf{F4} is a normalized LODF metric:
\begin{enumerate}
     \item 
     {\textit{Minimization of Firewall Resources and ACLs:}}
    The number of firewalls deployed within a network depends on the cyber topology and the number of clusters that require protection.
    Within each firewall, the number of ACLs depends on the number of network and port object-groups as well as the policies defined. The number of port object-groups are based on the service flows that traverse
    the network, while the network object-group depends on the number of sub-nets that access the protected network.  Hence, we define the objective function to minimize the firewall resources as $FS_{metric}$,
    \begin{equation}\label{fs_metric}
        FS_{metric} = N_f + N_{ACL}
        = \mathbf{F1} + \mathbf{F2}\\
    \end{equation}\label{eq:fs}
    \noindent where $N_f$ is the number of firewalls and $N_{ACL}$ is the number of ACLs. $N_f$ depends on the number of security zones $N_{sz}$, i.e., $N_{f} = f(N_{sz})$, while $N_{ACL}$ depends on the number of network and port object-groups, i.e., $N_{ACL} = g(N_{noj}, N_{poj})$. For a given set of service flows, $N_{poj}$ remains fixed, but $N_{noj}$ depends on the number of security zones, i.e., $N_{noj} = h(N_{sz})$. The functions $f,g,h$ are linear in $N_{sz}$ and are discussed in detail in the results.

    \item {\textit{Maximization of Physical Security Metric
    :}} The physical security metric is given by the least number of protection devices that can be jeopardized if compromised. Hence we define the metric as $PS_{metric}$,
    \begin{equation}\label{ps_metric}
        \mathbf{F3} = PS_{metric} = \sum_{w \in W, n \in N}{} \frac {1}{w*n}
    \end{equation}
    \noindent where $W=$ \{$w_{iso}, w_{cb}, w_{xline}, w_{xfmr}$\} and $n=$ \{$N_{iso}, N_{cb}, N_{xline}, N_{xfmr}$\}. The weights depend on various factors, e.g., the electrical equipment they connect.
    For instance, circuit breakers connecting to a generator with higher capacity would have higher weights, and 
    transformer 
    weights would depend on their power ratings.
    
    \item {\textit{Line Outage Distribution Factor (LODF):
    } LODFs find the transmission lines whose loss has the highest impact on the power flow in the network. A normalized LODF metric is computed as $NLODF$~\cite{lodf}. 
    \begin{equation}\label{eqn:LODF}
       \mathbf{F4} = NLODF = \frac {mean(abs(LODFs))}{std(abs(LODFs))}
    \end{equation}
    
    The goal of the clustering is to reduce the LODF associated within each network cluster.}
 \end{enumerate}

\subsection{Constraints} The following \textbf{constraints} are considered:
\begin{enumerate}
    \item {The number of clusters within a subgraph has an upper and lower limit, 
    \begin{equation}
    \small
g_{1} \equiv P^{\min } \leq P \leq P^{\max }
\end{equation}

where [$P^{\min},P^{\max}$] is the range of the number of clusters within a sub-graph.
    }
    \item {There exists a lower limit of the nodes within a cluster: 
    \begin{equation}
    \small
g_{2} \equiv n_{p}^{ \min } \leq n_{p}, \quad p=1 \text { to } P
\end{equation}
    }
    
    \item {There exists at least one node in a cluster that had a UCC node as its neighbor in the original graph.}
\end{enumerate}


%% file: Frontiers_Latex_Templates/Journal_SectionWise/4_Proposed_Solution.tex
\section{Proposed Meta-Heuristic Solution Approach}\label{solution}

A genetic algorithm \cite{538609} is a meta-heuristic algorithm based on natural selection, with the algorithm initiating a random set of possible solutions, called a \textit{population}  of solutions. These solutions are evaluated based on a set of fitness functions, where the individuals having the best adaptation measure have higher chances of reproducing and generating new offspring. 
The generation process consists of \textit{crossover} and/or \textit{mutation} operators, and it continues repeatedly until a global optimal solution is obtained. 
A crossover operator creates an offspring by combining parts of two parent solutions. 
There are single point and multi-point crossover. 
In our experiments, we have evaluated the computation time of the algorithms with varying number of crossover points. Increased number of crossover points results in higher disruptivity hence making the algorithm difficult to converge to an optimal solution. In certain cases, such as in later stages of the search process when the population is homogeneous or when the population size is small, larger crossover points can be beneficial.  
The mutation operator is used to maintain genetic diversity (i.e., variation) from one generation of population to the next. 
There are different types of mutation operators such as polynomial, inversion, binary bit flipping, etc. 
In this work, binary bit flip mutation is performed, where some selected bits within the chromosome are flipped based on a constant probability or the probability of $\frac{1}{dec\_var}$, where $dec\_var$ is the number of decision variables or length of the chromosome.

A single fitness/objective function cannot provide optimal network clusters; hence, a multi-objective genetic algorithm is adopted with the focus of increasing resiliency and reducing firewall resources. NSGA~\cite{996017} has been found to efficiently solve constrained multi-objective problems. Therefore, we adopt NSGA-2 to solve the optimal zoning problem. This technique performs well for a maximum of 3 to 4 fitness functions, and for more objective functions,  Multi-Objective Evolutionary Algorithms by Decomposition (MOEAD) or NSGA-3 can be explored.

\subsection{Algorithm Description}
The 
NSGA-2 algorithm (Alg.~\ref{alg:nsga2}) 
involves two main steps: \subsubsection {Offspring generation and sorting} \textbf{(a)} From the given population, $P_t$, at iteration $t$, the offspring solution, $Q_t$, is obtained using the selection, mutation, and crossover operations (Line 12-15). In the first step, using the union of $P_t$ and $Q_t$, non-dominated sorting is performed to obtain solutions at different pareto-front levels (Line 2-3). Non-dominated sorting in multi-objective problems is a sorting done between two solutions, say $X$ and $Y$, where $X$ is considered to dominate $Y$ if and only if there is no objective of $X$ worse than that objective of $Y$ and there is at least one objective of $X$ better than that objective of $Y$. Further, pareto-front of a multi-objective problem is a set of non-dominated solutions, which are chosen as optimal if no individual objective can be improved without sacrificing at least one other objective. 
\subsubsection {Obtain diverse solutions} \textbf{(b)} In the second step, while the next population set $P_{t+1}$ is obtained by sequentially adding the elements in the obtained pareto fronts, starting with 1 until the condition $|P_{t+1}| + |F_i| \le N$ is satisfied (where $F_i$ is the solution in the $i^{th}$ front, and $N$ is the maximum size of the population), for the selection of the elements in $F_i$, crowding-distance computation using the fitness function in each front (Line 6) is performed to obtain diverse solutions (Line 5-9).

\begin{algorithm}
\small
\caption{Pseudo-code for NSGA-2 \cite{996017}}\label{alg:nsga2}
\begin{algorithmic}[1]

\While{termination criteria}
\State $R_t \gets P_t \cup Q_t$
\State $F \gets $ non\_dominated\_sorting($R_t$)
\State $P_{t+1} \gets \phi ; i \gets 1$
\While {$|P_{t+1}| + |F_i| \le N$}
 \State $C_{i} \gets $ crowd\_sourcing\_assignment($F_i$)
 \State $P_{t+1} \gets P_t \cup F_i$
 \State $i = i + 1$
\EndWhile
\State $F_i \gets sort(F_i, C_i, desc)$
\State $P_{t+1} \gets P_{t+1} \cup F_i[1:(N - |P_{t+1}|)]$
\State $Q_{t+1} \gets selection(P_{t+1},N)$
\State $Q_{t+1} \gets mutation(Q_{t+1}) $
\State $Q_{t+1} \gets crossover(Q_{t+1})$
\State $t \gets t+1$
\EndWhile
\end{algorithmic}
\normalsize
\end{algorithm} 



\begin{table}[t]
\caption{NSGA-2 algorithm \& network clustering problem attributes}
\begin{adjustbox}{width=0.49\textwidth}
{\renewcommand{\arraystretch}{1.5}
\begin{tabular}{>{\centering}m{4cm}|>{\centering}m{1.4cm}| >{\centering}m{4cm}|>{\centering}m{1.4cm} } 

\hline
\hline
    \textbf{Problem Attributes} & \textbf{Ranges} &    \textbf{Algorithm Parameters} & \textbf{Ranges} \\   
\hline
Number of nodes & [2-48] & Population size & [50-200]\\ \hline
Number of edges &  & Max no. of generations & 100 \\ \hline
Number of objectives & [2-4] & Offspring generation & [10-50] \\ \hline
Number of constraints & 3 & Crossover points & [10-50]\\ \hline
Range of clusters & [2-40] & Crossover probability & 0.9\\ \hline
Minimum nodes within partition & 1 & Mutation probability & 0.05 \\ 
   \hline
   \hline
 \end{tabular}}
\end{adjustbox}
\label{configuration}
\end{table}

\subsection{Firewall Optimization Solution in Detail}\label{formulation}
Our solution is designed to find the optimal network cluster. The way this works is that when the network clusters are formed, the number of firewalls deployed in the $UCC$s and their substations along with the ACLs configured in those firewalls will reduce.   
Assuming a UCC and its associated substations as one graph, network cluster formation through edge elimination creates multiple sub-graphs. The solution is computed as a vector of subgraphs $x = \{1, \ldots, N_{sg}\}$, where for each solution, $N_{sg}$ is the total number of subgraphs that are formed.
The first $M_{u}$ subgraphs $x = \{1, \ldots, M_{u}\}$ are those that include a $UCC$ node, while subgraphs $x = \{M_{u}+1, \ldots, N_{sg}\}$ do not. The number of ACLs in the $UCC$ and substations are given by $ACL_{sub}$ and $ACL_{ucc}$,

\begin{equation}\label{acl_sub}
    ACL_{sub} = \sum_{x=1}^{M_u} 6(N_x - 1) + \sum_{x = M_u + 1}^{N_{sg}} 6
\end{equation}

\begin{equation}\label{acl_ucc}
    ACL_{ucc} = \sum_{x=1}^{M_u} 2(N_x - 1) + 5 + \sum_{x = M_u + 1}^{N_{sg}} 2
\end{equation}

\noindent where $N_x$ is the number of nodes in the subgraph $x$. In (\ref{acl_sub}), the constant \textit{6} is due to \textit{3} \textit{permit} rules for a) access to DNP3 outstations from DNP3 master, b) access to local database server, and c) access to local web-server; then there are \textit{3} \textit{deny} rules for
\texttt{ip\;any\;any}, associated with three interfaces of the firewall in the substation. In (\ref{acl_ucc}), the constant \textit{2} is due to 2 \textit{permit} rules that are a) \textit{SCADA server} in the $UCC$ access to the OT network, and b) \textit{HMI node} access to the $Local\_Web\_Server$, while the constant \textit{5} is due to \textit{3} fixed \texttt{ip\;any\;any} \textit{deny} rules associated with three interfaces of the firewall in the UCC connected to the substations end and \textit{2} fixed \textit{permit} rules for $SQL$ access from the $ICCP\_Server$ to $Public\_Database$ and $SCADA\_Server$ to $Public\_Database$. 

The net ACLs for a UCC is the sum of the ACLs for the UCC (\ref{acl_ucc}) and for its substations (\ref{acl_sub}), given by $N_{ACL}$:
\begin{equation}\label{eqn:nacl}
    N_{ACL} = ACL_{ucc} + ACL_{sub}
\end{equation}


The number of firewalls in a solution $x$ is given by $N_f$ (\ref{acl_sub_2}). In the first term, the subgraphs $x = \{1, \ldots, M_u\}$ include a $UCC$ node and are not clustered, so the number of firewalls is equal to the original number of nodes.  In the second term, the subgraphs $x = \{M_u + 1, \ldots, N_{sg}\}$ do not include the $UCC$ node, and there is one firewall for each subgraph:
\begin{equation}\label{acl_sub_2}
    N_f = \sum_{x=1}^{M_u} N_x - 1 + \sum_{x = M_u + 1}^{N_{sg}} 1
\end{equation}

The $N_{ACL}$ (\ref{eqn:nacl}) and  $N_f$~(\ref{acl_sub_2}) are computed and then used to update the fitness metric presented in (\ref{fs_metric}).

%% file: Frontiers_Latex_Templates/Journal_SectionWise/5_Abridged_Tool.tex
\section{Large-Scale Firewall Generation Tool \\ for Cyber-Physical Power Systems}\label{tool}

Deploying firewall policies for a large scale cyber-physical power system needed an application to automatically generate the configurations and also provide an interface to the users as a suggestion for re-configuration. 
The tool addresses, \textit{how can these optimal firewall rules can be automatically generated and implemented in a large-scale power system}? 
The research of proactive cyber-physical grid resilience techniques and tools 
depends on the firewall rules configured at different locations in the network.
However, manually configuring the firewall rules for a larger grid is a tedious task. The 
\emph{firewall generation} desktop application tool we have developed 
can be easily deployed to any network emulator or simulator to help translate the research to testbeds and then  industry.
Our prior work~\cite{TPECFW2022} elaborates the functionality and usage of the application in detail.
In this work, the tool is used to extract the existing firewall model and incorporate the results of the optimal network clustering to re-configure the policies that can be emulated in the RESLab testbed.

\subsection{Workflow of Firewall Configuration Tool}
\begin{figure}[h]
\centering
\includegraphics[width=0.9\linewidth]{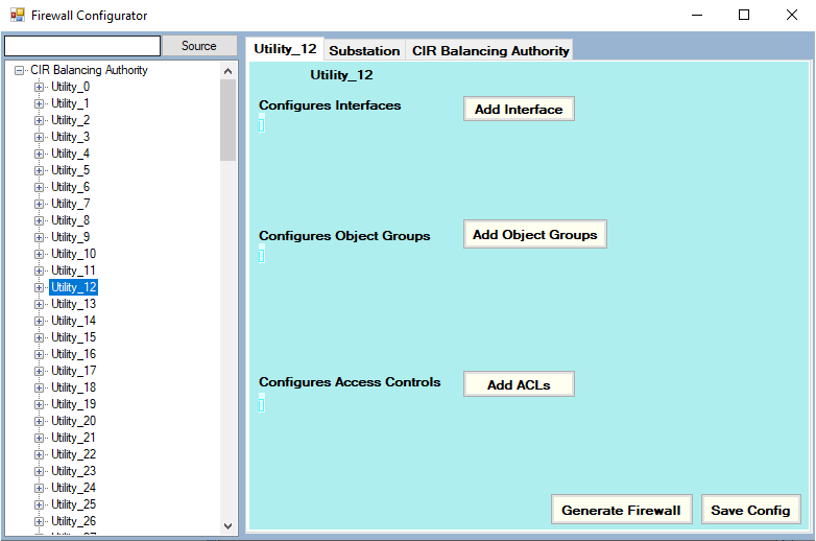}
\caption{Screenshot of the firewall configuration panel to configure interfaces, network and port object groups, and ACLs~\cite{TPECFW2022}. 
}
\label{fig:app_GUI}
\end{figure}

The workflow of the 
\emph{firewall generation} 
application is detailed in~\cite{TPECFW2022}
which is used to streamline the analysis in this paper. 
For the synthetic 2000-bus cyber network, Fig.~\ref{fig:app_GUI}  
lists utility control centers in a tree view, where users can configure interfaces, object groups and ACLs. 



\subsection{
Incorporation of Firewall Policy Analysis}

Based on the configuration of the interfaces, network and port object groups, and ACLs for the firewalls, the application provides a visualization of different networks. These interactive visualizations with firewall optimization analytics have broad application, such as monitoring the traffic that have been hit based on the firewall policy, within the
%
next-generation cyber-physical energy management systems.

\begin{figure*}[hbt!]
\centering
\includegraphics[width=0.8\linewidth]{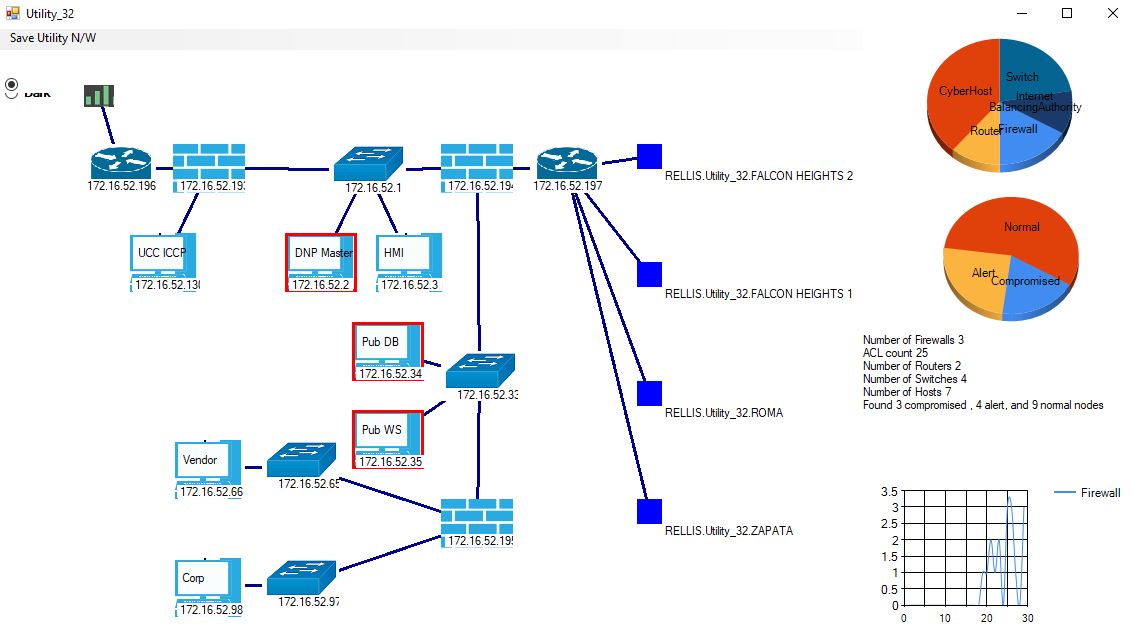}
  \caption{The Utility\_32 network's topology in the CYPRES application with the firewall generation tool that configures the rules~\cite{TPECFW2022}.}
  \label{fig:ucc_diag}
\end{figure*}

\begin{figure}[h!]
\centering
\includegraphics[width=1.0\linewidth]{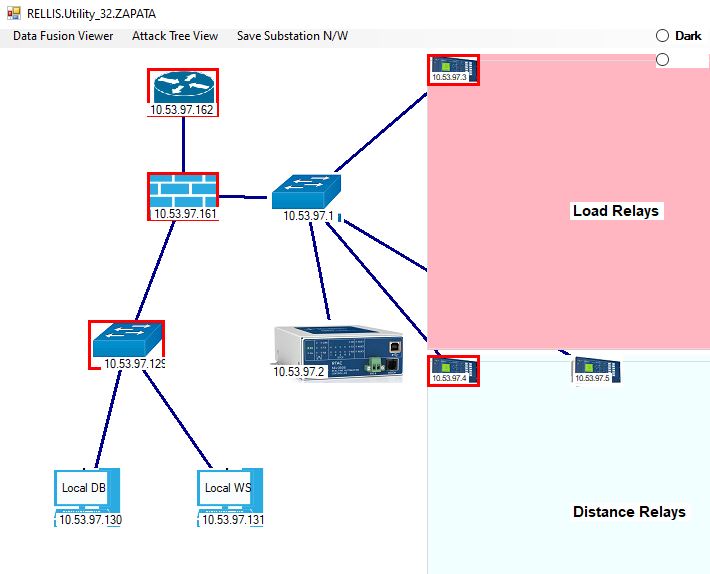}
  \caption{The Substation ZAPATA network's topology in the CYPRES application that interacts with the firewall tool to configure the rules. }
  \label{fig:sub_diag}
\end{figure}

An example (detailed further in~\cite{TPECFW2022}), is given of the UCC for Utility\_32 network topology, whose firewall rules are configured based on the firewall generation tool is shown in Fig.~\ref{fig:ucc_diag}. On the left side, there is the DMZ that connects the UCC to a BA, with its ICCP server. In the middle, at the top, there is the main SCADA  workstation, such as the DNP master and HMI machine. On the left, a firewall and a router connects the UCC to all substations belonging to this utility, a total of four substations. Other DMZs are shown at the bottom, such as the public, vendor, and corporate DMZ's.  For this Utility\_32 control center, there are three firewalls, two routers, four switches, seven hosts, and a total of 25 ACL rules generated. Fig.~\ref{fig:sub_diag} shows the Substation ZAPATA communicating the Utility\_32 control center through router with IP \texttt{10.53.97.162}.

For the complete 2000-bus synthetic model, assuming substations are connected in a star topology to the UCC, the firewall generation tool computed the number of firewalls ($N_f$), number of networks ($N_{noj}$) and port object groups ($N_{poj}$), and number of ACLs ($N_{acl}$). The result of all these metrics in shown in Table~\ref{stats}.

\begin{table}[b]
\caption{Cyber component and firewall configuration 
results for the 2000-bus synthetic grid model. 
}
\begin{center}
\renewcommand{\arraystretch}{1.2}
\begin{tabular}{||c|c||}
\hline
\textbf{Components} & \textbf{Counts}\\
\hline
\hline
Total Number of nodes & 18463\\
\hline
Total Number of Firewalls & 1702\\
\hline
Firewalls in UCCs & 450\\
\hline
Network Object Groups (UCCs) & 1200 \\
\hline
Port Object Groups (UCCs)& 4800 \\
\hline
Access Control Lists (UCCs) & 5050\\
\hline
Firewalls in Substations & 1250 \\
\hline
Network Object Groups (Substations) & 6250 \\
\hline
Port Object Groups (Substations) & 3750 \\
\hline
Access Control Lists (Substations) & 10000\\
\hline
\end{tabular}
\label{stats}
\end{center}
\end{table}

%% file: Frontiers_Latex_Templates/Journal_SectionWise/6_Results_Analysis.tex

\section{Results and Analysis}\label{results}
In this section, we 
evaluate the pareto-optimal solution obtained using the NSGA-2 algorithm, on the basis of the fitness functions. We analyze the impact of the genetic algorithm parameters, detailed in Table~\ref{configuration}. Further, we study the impact of selection of the fitness function, followed by comparison of the star, hybrid, and optimally-clustered hybrid topologies for the synthetic 2000 bus case.

\subsection{Interpretation of NSGA-2 solution}

Our approach forms optimal network clusters through consideration of both cost and resilience. 

To improve cost, the objective is to segment the network to reduce the firewalls deployed (\ref{acl_sub_2}) and the net amount of ACLs configured (\ref{eqn:nacl}).
Costs are important to include in the network segmentation problem. A star topology requires significant
cybersecurity investment by a utility
since it requires deployment of more firewalls, with a higher computation burden on the processors due to the huge list of ACL rules as shown 
in Table~\ref{stats}. Usage of a hybrid topology (e.g., substations connected to a UCC and to each other) without any network clusters also makes the cybersecurity investment high.

To improve resilience, the objective is to segment the network to reduce the risk encountered by deploying greater numbers of critical equipment in a single cluster. The risk is minimized by increasing the number of clusters, while reducing the LODF metric of each cluster, as described in Section~\ref{problem}.
%
Increasing the number of network clusters ensures that the critical assets are distributed more uniformly and reduces both the access of intruders to cause damage and the extent of possible damage.
Reducing the LODF metric in the segmentation also results in ensuring uniform distribution of critical transmission line with high LODFs across multiple network clusters. 

A set of optimal solutions obtained for a utility with 37 substations is given in Table~\ref{samplesoln}, which shows six sample solutions ranging from 6 to 21. Results indicate that a higher \# of clusters ensures resilience, based on rise and fall in trends for $F3$ and $F4$ respectively, but at the cost of increase in cost, i.e., rise in $F1$ and $F2$. The exception of the trend is observed for the case with 18 clusters, with respect to $F1$ and $F2$, may be due to existence of more \# of subgraphs without $UCC$ node. Similarly, an exception is observed in trend for the case with 6 clusters, with respect to $F4$, hence it may be preferable to select solution with $6$ instead of $21$, if LODF is a prefferred criteria over overall security index $F3$.

\begin{figure*}[h!]
  \centering
  \subfigure[]{\includegraphics[width=0.195\linewidth]{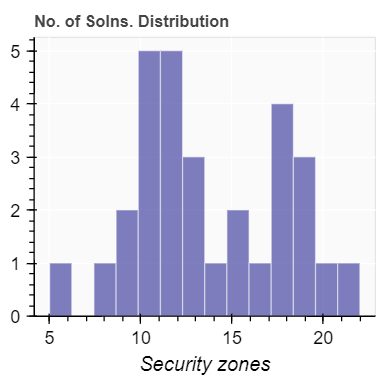}\label{sol_dist}}
  \subfigure[]{\includegraphics[width=0.195\linewidth]{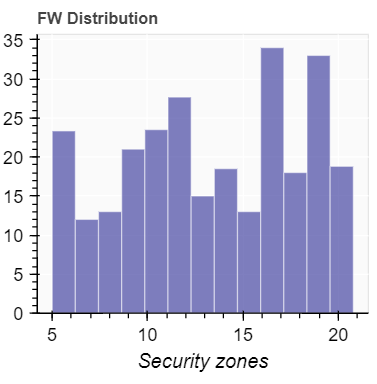}\label{f1_effect}}
  \subfigure[]{\includegraphics[width=0.195\linewidth]{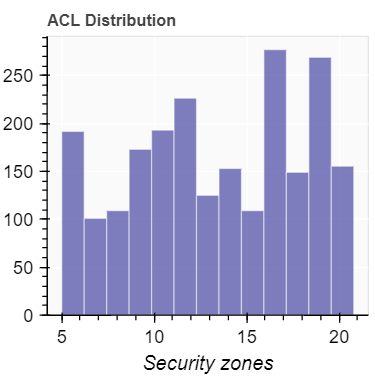}\label{f2_effect}}
  \subfigure[]{\includegraphics[width=0.195\linewidth]{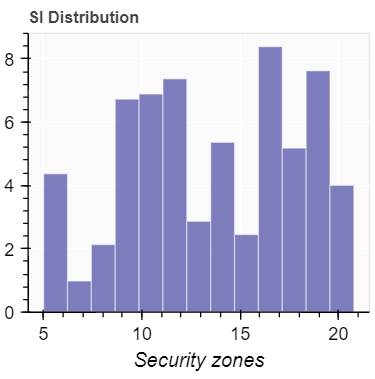}\label{f3_effect}}
  \subfigure[]{\includegraphics[width=0.195\linewidth]{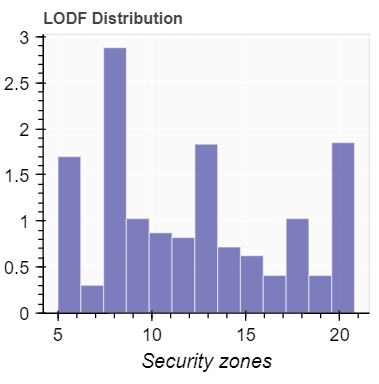}\label{f4_effect}}
  \caption{Security zones distribution along with average fitness scores.}
\end{figure*}

The Table~\ref{samplesoln} analyzes 6 solutions from a total of 30 pareto-optimal solutions obtained. 
Fig.~\ref{sol_dist} shows the complete results with the network cluster distribution (for instance, there are 40 solutions with cluster less than 3).
While the Figs.~\ref{f1_effect},~\ref{f2_effect},~\ref{f3_effect},~\ref{f4_effect} shows the average fitness scores, \textbf{F1} to \textbf{F4}, respectively, across the range of network clusters.  Based on the distribution, we deduce that solution with 12 to 16 \# of clusters are more cost-effective, while 16 to 20 clusters, can be more resilient given higher SI and lower LODF.


\begin{table}[hbt!]
\caption{A set of optimal solutions for a utility with 37 substations obtained using NSGA-2 algorithm. \textbf{F1}: number of firewalls, \textbf{F2}: number of ACLs, \textbf{F3}: Security Index, and \textbf{F4}: LODF metric }
\begin{center}
\renewcommand{\arraystretch}{1.2}
\begin{tabular}{||c|c|c|c|c||}
\hline
\textbf{No of Network clusters} & \textbf{F1}& \textbf{F2} & \textbf{F3} & \textbf{F4}\\
\hline
\hline
6 & 12 & 101 & .98 & 0.3\\
\hline
9 & 14 & 117 & 2.88 & 1.23\\
\hline
12 & 16 & 133 & 4.21 & 1.03\\
\hline
15 & 20 & 165 & 5.68 & 1.03\\
\hline
18 & 18 & 149 & 6.87 & 0.41 \\
\hline
21 & 34 & 277 & 8.38 & 0.41\\
\hline
\end{tabular}
\label{samplesoln}
\end{center}
\end{table}


\subsection{Effect of Genetic Algorithm Parameters}
\begin{figure*}[h!]
  \centering
  \subfigure[]{\includegraphics[width=0.32\linewidth]{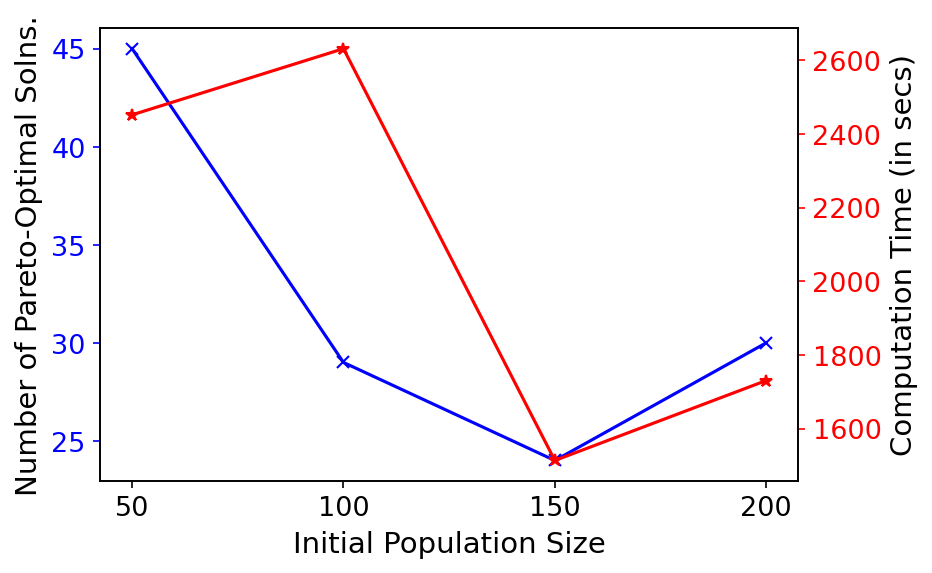}\label{ps_effect}}
  \subfigure[]{\includegraphics[width=0.32\linewidth]{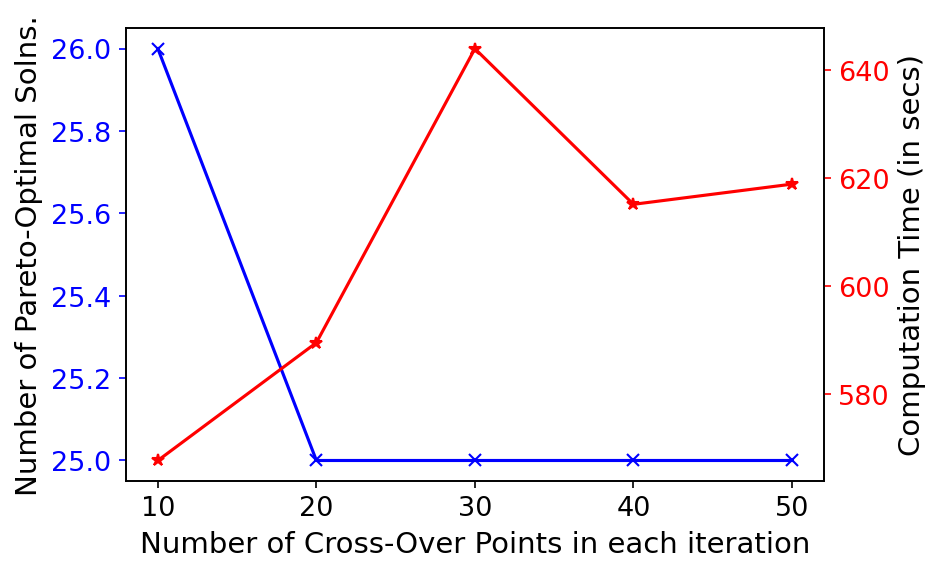}\label{cp_effect}}
  \subfigure[]{\includegraphics[width=0.32\linewidth]{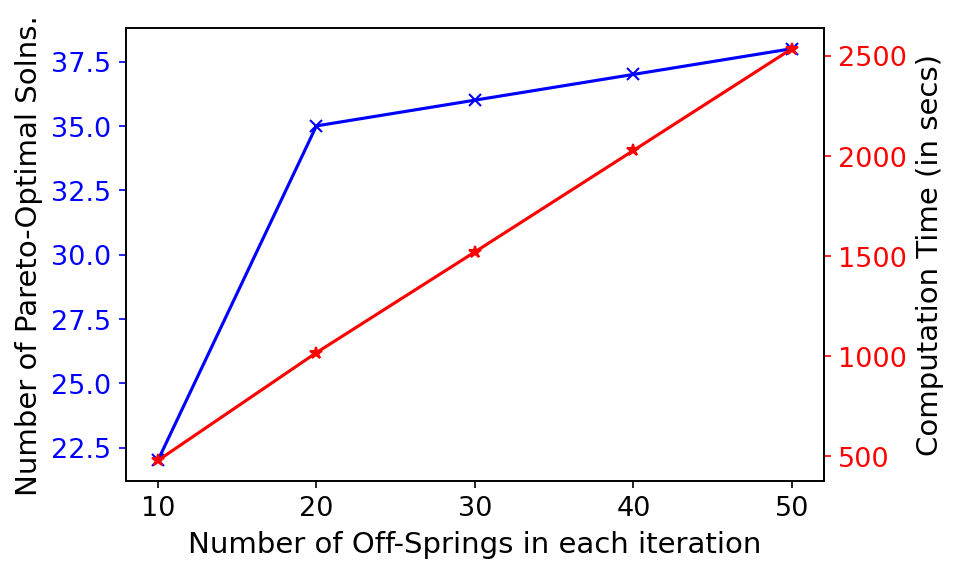}\label{os_effect}}
  \caption{Effect of (a) population size (tested for a utility with 36 substations), (b) number of crossover points (tested for a utility with 10 substations), and (c) number of offspring in each generation (tested for a utility with 12 substations).}
\end{figure*}

Identification of the ideal parameters of the NSGA-2 algorithm is essential. These are based on providing pareto-optimal solutions or computation time, and they are unique and problem dependent.
Three common GA parameters are considered: a) population size for each generation, b) crossover points in mutation, and c) offspring size in every generation. 
Figs.~\ref{ps_effect},~\ref{cp_effect}, and ~\ref{os_effect} shows the effect of the parameters of the NSGA-2 algorithm.
The time complexity of the NSGA-2 algorithm is $O(MN^{2})$, where $M$ is the number of objective functions, and $N$ is the initial population size $ps$ followed by the offspring generated in every iteration. Fig.~\ref{os_effect} illustrates the increase in computation time with increasing offspring size, while Figs.~\ref{ps_effect} shows that computation time reduces with a higher initial population size, but reduces the number of solutions obtained, e.g., an initial population of 150 results in 25 solutions. Based on the computation time and the number of solutions that increases with a $ps$ of 200, $ps = 150$ 
is considered for all the simulations.
The computation time was not much affected by number of crossover points (Fig.~\ref{cp_effect}). However, the algorithm can be sensitive to an increased number of crossover points, making it difficult to converge to an optimal solution; however, in certain cases---such as in the later stages of the genetic algorithm search process, when the population is homogeneous, or for a smaller population size---a larger number of crossover points can be beneficial~\cite{cp_analysis}.

\subsection{Effect of Selection of Fitness Function}
Since NSGA-2 solves the multi-objective problem, in this section, we evaluate the impact of selection of subset of fitness functions on the optimal solutions. As the \# of fitness function increases, depending on the \# of resilience or cost metric considered, the performance of the algorithm deteriorate~\cite{996017}. Hence, it is essential to perform the analysis of finding optimal solutions, by selecting a subset of fitness functions.
The experiments are performed by taking the combination of two and three objective function from the four described in the Section~\ref{problem}. 
The results on the computation time, number of optimal solutions, and the average number of clusters obtained for varying objective functions are shown in Table~\ref{varyobj}. It can be observed that using the physical objective functions \textbf{F3} and \textbf{F4} increases the network clusters, while the usage of cyber objective functions \textbf{F1} and \textbf{F2}, has the tendency to reduce the clusters so as to curtail the budget. The \# of pareto optimal solutions decreases with increase of fitness function, with some exception of $F3,F4$ with very few solutions as 3. 

\begin{table}[hbt!]
\caption{Solutions under varying objective functions.}
\begin{center}
\renewcommand{\arraystretch}{1.2}
\begin{tabular}{||c|c|c|c||}
\hline
\textbf{Obj Funcs.} & \textbf{Comp Time}& \textbf{Solns} & \textbf{Avg. NW Clusters}\\
\hline
\hline
F1,F2 & 1476 & 78 & 3\\
\hline
F1,F3 & 1472.78 & 200 & 4.4\\
\hline
F1,F4 & 1477.2 & 141 & 3.5\\
\hline
F2,F3 & 1477 & 200 & 4.4\\
\hline
F2,F4 & 1479.5 & 141 & 3.4\\
\hline
F3,F4 & 1482.4 & 3 & 12\\
\hline
\hline
F1,F2,F3 & 1281.3 & 26 & 6.31\\
\hline
F1,F2,F4 & 1315 & 3 & 3.33\\
\hline
F2,F3,F4 & 1404 & 26 & 6.58\\
\hline
F1,F3,F4 & 1151.6 & 26 & 6.58\\
\hline
\end{tabular}
\label{varyobj}
\end{center}
\end{table}

\subsection{Evaluation of Computation Time
vs.
Sub-Graph Size}
\begin{figure}[h!]
\centering
\includegraphics[width=0.81\linewidth]{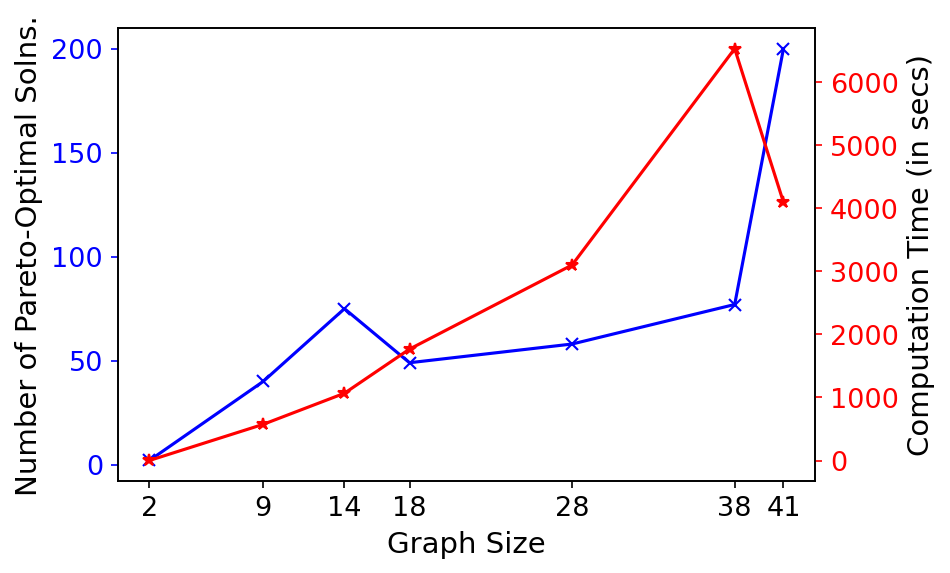}
  \caption{Evaluation of computation type for different NSGA-2 problem of varying graph size}
  \label{fig:computation}
\end{figure}

Fig.~\ref{fig:computation} shows the evaluation of computation time for varying graph size i.e. graph generated for a utility control center with their substations. The number of decision variables (i.e. the edges to eliminate), modeled for segmentation increases with the size of graphs, hence the mutation, cross-over and selection operations increases with the number of $dec\_var$.

\subsection{Statistics of the Pareto Optimal Solution}
\begin{figure*}[h!]
  \centering
  \subfigure[]{\includegraphics[width=0.245\linewidth]{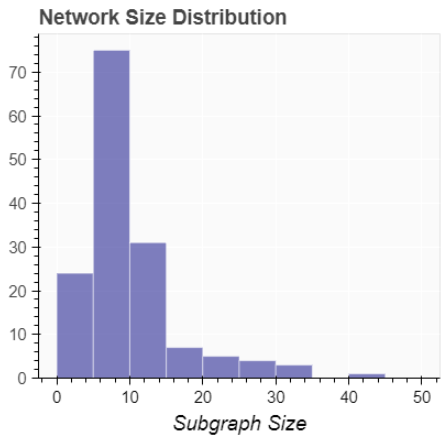}\label{gs_dist}}
  \subfigure[]{\includegraphics[width=0.245\linewidth]{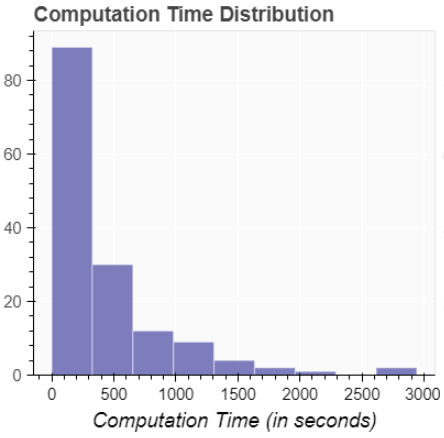}\label{ct_dist}}
  \subfigure[]{\includegraphics[width=0.245\linewidth]{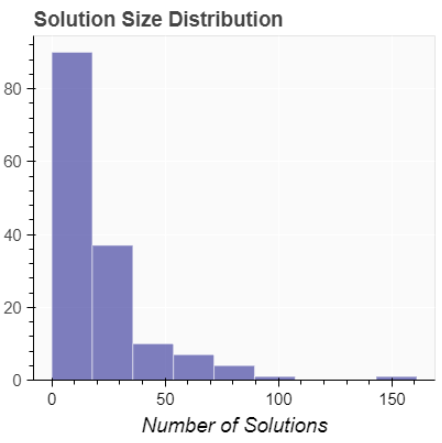}\label{soln_dist}}
  \subfigure[]{\includegraphics[width=0.245\linewidth]{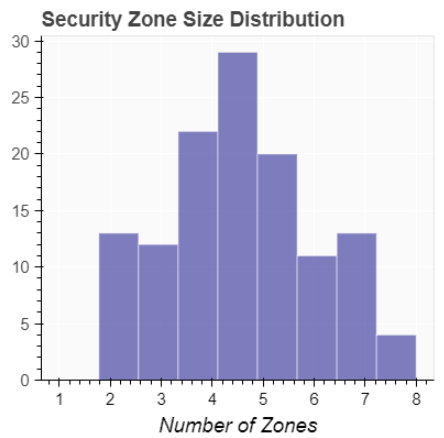}\label{sz_dist}}
  \subfigure[]{\includegraphics[width=0.245\linewidth]{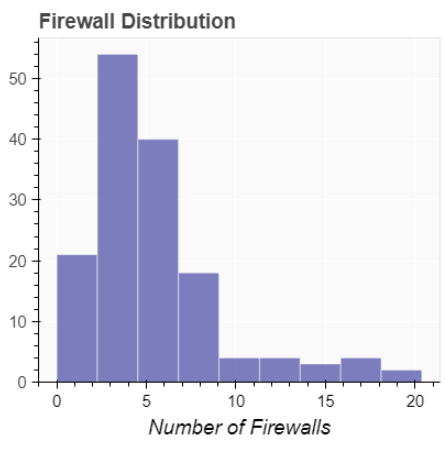}\label{f1_dist}}
  \subfigure[]{\includegraphics[width=0.245\linewidth]{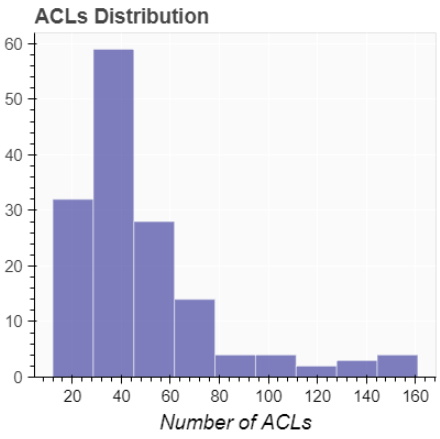}\label{f2_dist}}
  \subfigure[]{\includegraphics[width=0.245\linewidth]{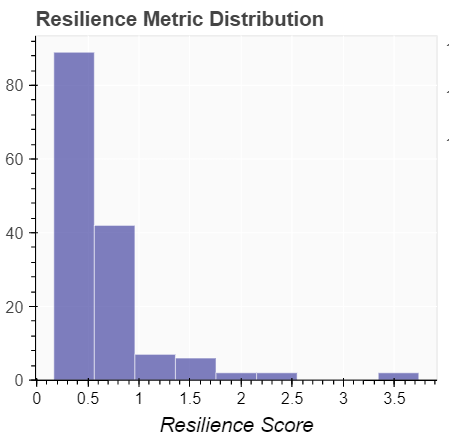}\label{f3_dist}}
  \subfigure[]{\includegraphics[width=0.245\linewidth]{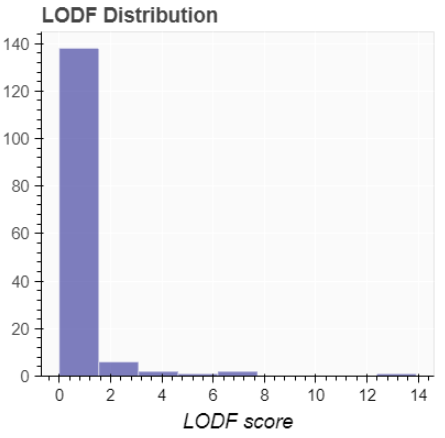}\label{f4_dist}}
  \caption{Distribution of the (a) Subgraph Size, (b) Computation Time, (c) \# of Optimal Solutions, (d) \# of Network Clusters and the fitness functions: (e) \# of Firewalls (F1), (f) \# of ACLs (F2), (g)Physical Security or Resilience Metric (F3), (h) LODF metric (F4), considered in the NSGA-2 algorithm for the synthetic 2000 bus case with 1250 substations and 150 UCC.}
\end{figure*}

\begin{figure}[h!]
\centering
\includegraphics[width=1.0\linewidth]{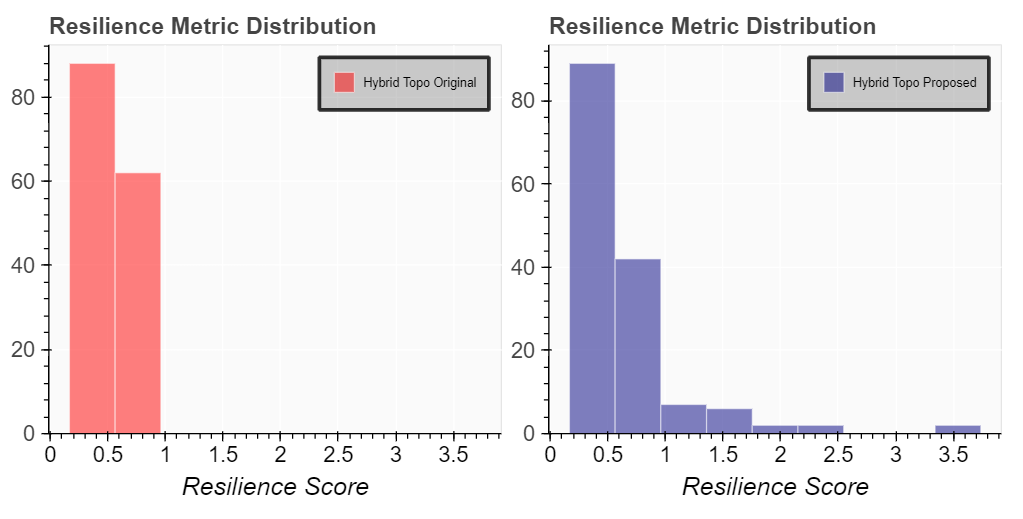}
  \caption{Comparison of $PS_{metric}$ distribution for hybrid topology original and proposed solution. 
  }
  \label{fig:compare_si}
\end{figure}

As discussed earlier, 150 utilities along with their substations are modeled in the synthetic communication network. For each utility along with its substation, a graph is modelled, which are optimally segregated based on solving 150 different NSGA-2 algorithms. Fig.14 (a-h) shows the distribution of those 150 graphs based on the network size, computation time for solving the optimization problem, \# of pareto-optimal solutions, \# of network clusters and the scores of the four fitness functions.  The x-axis in the plot refers to the absolute values of the 8 different metrics in the x-axis labels, while y-axis represent the frequency of occurrence in the given range. For instance, from Fig.~\ref{gs_dist}, there are 74 UCCs whose network size ranged from 5-10 nodes and from Fig.~\ref{f1_dist}, there are 54 UCCs whose number of firewalls are around 3 to 5. 
Such evaluation can assist in estimating the budgets and risk involved in the proposed solution. For instance, from Table~\ref{stats} the total \# of firewalls in the star topology is 1250 and hybrid topology without clustering is 2267 (\# of firewalls in each substation, is equal to the degree of the node) while from Fig.~\ref{f1_dist}, the total \# of firewalls in the hybrid topology with optimal clusters is the integration of Fig.~\ref{f1_dist} i.e. roughly 858, which is almost \textbf{31.5\%} and \textbf{62.2\%} reduction from star and hybrid topology respectively. Similarly, there were almost 11050 ACLs (10000 in substation, 1050 in UCC interfaced to substations) in star-topology, and 15710 ACLs in hybrid topology (computed based on the \# of firewalls in hybrid topology and ACLs based on Eq.~\ref{acl_sub} and \ref{acl_ucc}), while from Fig.~\ref{f2_dist}, there are almost 6580 ACLs, with almost \textbf{40.5\%} and \textbf{58.2\%} reduction of ACLs from star and hybrid topology respectively. The resilience metric too improved for some substations with the proposed solution as shown in the distribution comparison Fig.~\ref{fig:compare_si}.


%% file: Frontiers_Latex_Templates/Journal_SectionWise/7_Discussion.tex
\section{Discussions}\label{discussion}

In this paper, 
we extend~\cite{TPECFW2022} that presented an automatic firewall configuration and generation tool, by focusing here on how to optimally generate those configurations; the tool introduced in~\cite{TPECFW2022} is then applied in the experiments to implement and evaluate/validate the optimal designs in the RESLab testbed.
Additionally, \cite{TPECFW2022} designed networks using only a \textit{star} topology, while
here we also consider
\textit{hybrid} network topologies, with more electronic perimeters to secure and increased amount of security infrastructure (e.g, numbers of firewall devices, ACL rules, and security zones).
%
To address the higher 
cost associated with 
more electronic security perimeters, we proposed a novel optimal network security zone formation problem that can keep a balance between the cyber investment and grid resilience. 
Since it is a multi-objective optimization problem, we solved the problem using a meta-heuristic algorithm, NSGA-2, and evaluated the performance based on the genetic algorithm parameters, selection of fitness function, and computation complexity for varying sub-graphs formed for each UCC and their associated substations. 

%% file: Frontiers_Latex_Templates/Journal_SectionWise/8_Conclusion.tex
\section{Conclusion}\label{conclusion}

This work presented an optimization approach and associated modeling tools to design and configure optimal
firewall rules for electric power utilities
that ensure connections are only allowed based on the applications and services needed to operate, following  NERC-CIP-005 that requires an \emph{electronic perimeter} to secure utilities and their field devices.
A mathematical model was presented that computed the optimal network clusters to minimize the cyber investment cost to keep the grid more resilient.
The firewalls are tested in RESLab with the developed
automated firewall generation tool 
that would also allow other researchers and utility operators to implement 
firewall rules in a large-scale emulation network model for further validation and then transition to practice to real systems.
The optimization technique and the  firewall generation tool together allows users to find and configure optimal firewall policies in a scalable way, and to visualize the network topology and its firewalls in context the complete cyber-physical power system. 